\documentclass[prl, twocolumn, showpacs]{revtex4}
\usepackage{amssymb}
\usepackage{amsmath}
\usepackage{epsfig}
\usepackage{bm}
\usepackage{color}

\begin{document}

\title{ A classical   mechanism  for negative magnetoresistance
\\
in two-dimensional systems   in the  ballistic regime }
\author{P. S. Alekseev and M. A. Semina }
\affiliation{ Ioffe  Institute,  194021  St.~Petersburg, Russia }

\begin{abstract}
In  ultra-high quality two-dimensional (2D) materials
 the mean free paths of phonons and electrons relative
  to all mechanisms of scattering can be much
 greater than a size of a sample. In this case the most
  intensive type of scattering of particles is their collisions
   with sample edges and
   the ballistic regime of heat and charge transport is realized.
  We study the ballistic  transport
    of classical interacting 2D particles in  a long narrow sample.
 We show that the inter-particle scattering conserving momentum leads to
a positive {\em hydrodynamic} correction to the ballistic conductance,
 which is a precursor of the viscous Poiseuille flow.
 We examine  the effect of weak
 magnetic field on the electron ballistic conductance
and predict a novel classical   {\em ballistic } mechanism  for negative magnetoresistance.
 Our analysis demonstrates that,  apparently,  such mechanism explains
  the temperature-independent part of the giant negative magnetoresistance
 recently observed  in the ultra-high mobility
  GaAs quantum wells.
 \pacs{72.20.-i,    73.63.Hs,  72.80.Vp, 73.43.Qt }

\end{abstract}

\maketitle

{\em 1. Introduction. }
In solids with enough weak disorder the electron and phonon
mean free paths at low temperatures can be very large.
 In such case, charge and heat transport are realized
 by the ballistic mechanism in the narrowest samples
 or by the hydrodynamic mechanism in the samples of medium widths \cite{Gurzhi_rev,Gurevich_book}.
  In recent years,   the profound evidences
  of the hydrodynamic regime of charge transport were discovered
  in novel ultra-pure  materials:    high-mobility GaAs quantum wells,
 single-layer suspended graphene, 3D Weyl semimetals
   \cite{exp_hydgr_1,exp_hydgr_1_2,exp_hydgr_1_3,exp_hydgr_1_4,exp_hydgr_1_ov,exp_hydgr_2,exp_hydgr_3,exp_hydgr_4,exp_hydgr_4_2}.
 The experimental studies of hydrodynamic transport was accompanied by
 an extensive development of theory  \cite{hydr_tr_th_1,hydr_tr_th_2,hydr_tr_th_3,hydr_tr_th_4,hydr_tr_th_5,hydr_tr_th_6,hydr_tr_th_7,hydr_tr_th_8,hydr_tr_th_9,hydr_tr_th_10,hydr_tr_th_10_2,hydr_tr_th_11,hydr_tr_th_11_2,hydr_tr_th_11_3,hydr_tr_th_12,we_hd,we_hd_2,we_hd_3,we_hd_4,we_hd_5}.

One of the most bright  evidences of hydrodynamic regime of transport
 is the giant negative magnetoresistence effect observed
 in the high-mobility GaAs quantum wells and  the  3D Weyl semimetal WP$_2$ \cite{exp_hydgr_1,exp_hydgr_1_2,exp_hydgr_1_3,exp_hydgr_1_4,exp_hydgr_1_ov,exp_hydgr_2,exp_hydgr_3}.
 In Ref.~\cite{exp_hydgr_1_4} the term ``colossal negative magnetoresistance''
was even coined. This effect seemed outstanding and
 mysterious as the usual bulk theories of magnetotransport
 led to a positive magnetoresistance or to
  a moderate negative magnetoresistance in very weak magnetic fields
  (the weak localization effect). Recently the temperature-dependent part
   of the giant negative magnetoresistance
  was explained
 as the result of forming the viscous electron fluid
  and the magnetic field dependence of the electron viscosity \cite{hydr_tr_th_9}.
However, the temperature-independent part of the giant negative magnetoresistance,
 which occurs in some (but not in all) samples, has so far remained strange and unexplained.

Heat transport in single-layer suspended graphene
 had been extensively studied in recent several years
\cite{exp_gr_1,exp_gr_2,exp_gr_3,exp_gr_4}.
The measured values of the thermal conductivity coefficient
 are extremely high and dependent on the sample size.
 This indicates that heat transport in suspended  graphene is
  realized not by some bulk mechanism, related
  to the scattering of the flexural phonons
    on disorder or the phonon-phonon Umklapp scattering \cite{th_therm_gr_bulk_1,th_therm_gr_bulk_2,th_therm_gr_bulk_3,th_therm_gr_bulk_4,th_therm_gr_bulk_4_liftimes},
  but by the ballistic or the hydrodynamic mechanisms.
The anomalously large thermal conductivity of the suspended graphene
 is of huge interest in connection with the hope of using such samples in electronics.

 The ballistic mechanism of thermal conductivity in graphene was studied in
 Refs.~\cite{th_therm_gr_bal_1,th_therm_gr_bal_2}. The case of a short sample without
 any scattering of particles inside the sample was studied within the  Landauer approach \cite{th_therm_gr_bal_1}.
  The effect of the type of the scattering of the flexural phonon
  on sample edges, diffusive on rough edges
or specular on smooth edges, on the thermal conductance was examined \cite{th_therm_gr_bal_2}.
Hydrodynamic phonon transport in suspended graphene
 was recently  investigated in Refs.~\cite{th_therm_gr_hydr_1,th_therm_gr_hydr_2,paper_mail}.
In Refs.~\cite{th_therm_gr_hydr_1,th_therm_gr_hydr_2}
a numerical approach   was used  to demonstrate the  Pouseuelle flow
 of a phonon fluid. A rigorous analytical theory of hydrodynamics
 of the flexural phonons  was developed in Ref.~\cite{paper_mail}.

In this Letter we develop  an analytical theory of ballistic transport
 of {\em interacting } 2D particles in  long samples with rough boundaries.
 Our theory  is applicable to phonon and  electron heat transport
  as well as to charge transport in  graphene and
   quantum wells.   The main part of
  the free particle ballistic conductance of a long sample diverges as
  a logarithm of the sample width  \cite{log}.
  We show that this divergence is limited
    by the inter-particle collisions or by the finite sample length.
    The last case is, apparently, realized in the experiment \cite{exp_gr_4},
    manifesting itself by  the length  dependence
    of thermal conductivity of graphene samples.
 We demonstrate that the inter-particle collisions conserving momentum
   induces a {\em precursor } of the inhomogeneous viscous Poiseuille flow.

  For the electronic mechanism of heat and charge transport
   we study the
 effect of weak magnetic field on the ballistic conductance.
 The calculated magnetic field correction to the conductance
 is positive and  quadratic by magnetic field.
 It is related to the effect of magnetic field
 on free electron trajectories.
 Hereby, we propose a novel classical {\em kinematic } mechanism
 for negative magnetoresistance. Apparently,
 such magnetoresistance was observed in Ref.~\cite{exp_hydgr_1,exp_hydgr_1_2,exp_hydgr_1_3}
 as a temperature-independent  peak on the negative magnetoresistance
 curves against the background  of the large temperature-dependent peak.
  The last was
 explained in Ref.~\cite{hydr_tr_th_9} as a manifestation of
 viscous magnetotransport.
 Thereby  this work together with Ref.~\cite{hydr_tr_th_9} provides
 a complete explanation of the giant negative magnetoresistance effect observed
 in  the best-quality GaAs quantum-wells.

\begin{figure}[t!]
\centerline{\includegraphics[width=0.8\linewidth]{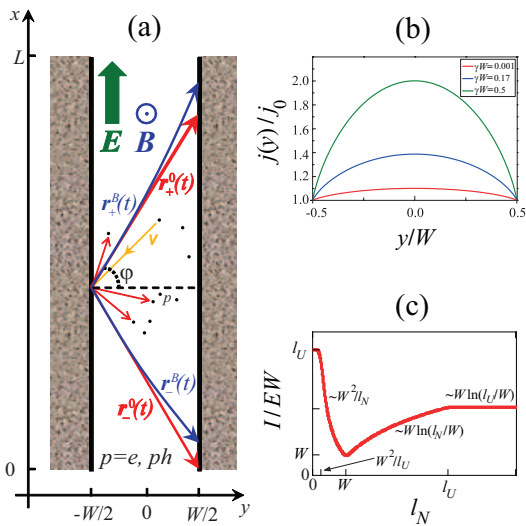}}
\caption{
(a)
A ballistic sample with particles $p$  and rough edges
 in external electric  and  magnetic fields $E$ and  $B$.
  Trajectories of charged particles ($p=e$),
  symmetrical  relative to the $y$ axis,
  are drawn for zero (red) and nonzero  (blue) magnetic fields.
  (b)
  The current density $j(y)$ at  different values
of the parameter $\gamma W $.
 (c)
 The current $I$ as a function
of the normal collision scattering length   $l_N $
 at a fixed value of the Umklapp  collision scattering length
   $l_U \gg W$.
 }
\end{figure}

{\em 2.  Model. }
 We consider a flow of 2D particles (phonons or electrons)
 in the sample  shaped as a long rectangular with the width $W$
 and the length $L \gg W$ [see Fig.~1(a)]. We study a linear response
 on a generalized  stationary external field $ E $
 which is  proportional to a gradient of temperature for
 the problem of  heat transport and  coincide with an external
 electric field for the problem of charge transport.
If the mean free paths relative to the inter-particle collisions
conserving and not conserving momentum, $l_N$ and $l_U$,
 are much larger than the sample width $W$,
 the collisions with  the longitudinal  sample
 edges are the most frequent type of scattering.

Further consideration is independent of
the particular type of particles, their dispersion laws
and the type of an external field. It is applicable to both heat
 and charge transport in  high-quality samples
 of graphene and other 2D materials. In this connection,
 for all quantities we use the units in which
  the characteristic microscopic quasiparticle
  velocity is equal to unity and coordinates, time,
   and reciprocal field, $1/E$, have the same units.

We assume the external field $E$ being small enough to retain
 particles in a state close to the thermal equilibrium.
We describe the particle response on $E$  by the
inequilibrium part of the distribution  function,
 $f(y, \varphi) \sim E $, in which the dependence on $x$
 is absent due to the relation $L \gg W$
 and  the energy dependence is omitted for simplicity.
 This approximation corresponds to neglecting
 the effect of the energy transfer during scattering events.

 The kinetic equation for the distribution function
 $f(y , \varphi)  $ has the form [see Fig.~1(a)]:
\begin{equation}
 \label{kin_eq}
\cos\varphi \, \frac{\partial f}{ \partial y }
 - \sin \varphi \, E
 = \mathrm{St}_N [f] + \mathrm{St}_U [f] \:,
\end{equation}
where  the collision integrals $\mathrm{St}_N $ and
$ \mathrm{St}_U $ describe the scattering mechanisms conserving
 and not conserving momentum. We follow
  Ref.~\cite{book__gas__the_method} and
  Refs.~\cite{hydr_tr_th_10,hydr_tr_th_10_2,hydr_tr_th_11,hydr_tr_th_11_2,hydr_tr_th_11_3}
in choice of the simplified forms of $\mathrm{St}_N $
and $ \mathrm{St}_U $:
\begin{equation}
 \label{St_N__St_U}
\mathrm{St}_N [f]  = - \gamma_N (f-P [f])
   , \;\;
\mathrm{St}_U [f]  = - \gamma_U (f-P_0 [f])  ,
\end{equation}
 where $\gamma_N$ and $\gamma_U$ are the corresponding
 scattering rates \cite{rates},  while  $P$ and $P_0$ are
 the projectors of $f(\varphi)$  on the subspaces consisting
 of the basis functions  $\{1,e^{\pm i \varphi} \}$ and $\{1\}$.
 The operator $\mathrm{St}_N$ conserves the perturbations
 of the distribution function corresponding to a nonzero current
 density and non-equilibrium concentration, while
the operator $\mathrm{St}_{U}$ conserves only the perturbations
of concentration.  For phonon transport $\mathrm{St}_N $
and $\mathrm{St}_U $ are related to the normal and
the Umklapp phonon-phonon collisions. For electron transport
$\mathrm{St}_N $ and $\mathrm{St}_U $  describe the electron-electron
scattering  and the electron scattering  on disorder or
  the electron-phonon scattering.

We assume the longitudinal sample edges being rough and
 the scattering of particles on them being fully diffusive.
Thus the boundary conditions on the distribution function are:
 $ f(-W/2, \varphi)=const_1 $ on the interval
 $ - \pi/2 < \varphi < \pi/2$  at the left edge
   and $ f( W/2, \varphi)=const_2 $
  on the interval  $  \pi/2 < \varphi < 3 \pi/2$
at the left edge [see Fig.~1(a)].

 The kinetic equation can be rewritten as:
\begin{equation}
\label{kin_eq_with_gamma}
\Big[\cos\varphi \, \frac{\partial }{ \partial y }
  + \gamma \Big] f- \sin\varphi \,E
 = (\gamma_N P  + \gamma_U P_0 )[f]
  \:,
\end{equation}
where $\gamma = \gamma_N +  \gamma_U  $ is the total scattering rate.
 In Ref.~\cite{SM} we show that in the hydrodynamic regime, $\gamma W \gg 1$,
  the left and the right parts of Eq.~(\ref{kin_eq_with_gamma})
 are of the same order of magnitude and Eq.~(\ref{kin_eq_with_gamma})
  transforms into the Navier-Stocks equation
 for the density of the flow of heat or charge $j(y)$.
  In the ballistic regime, $\gamma W \ll 1$,
 the terms in the left part of  Eq.~(\ref{kin_eq_with_gamma})
 are  much greater than the right part,  therefore  solving
 of Eq.~(\ref{kin_eq_with_gamma}) should
 be performed by the perturbation theory
 by the right part.

 We also demonstrate in Ref.~\cite{SM} that
 a perturbation of the particle density is absent in the considered problem:
 $ \delta n(y) = \int _0^{2\pi} d \varphi \, f(\varphi , y ) =0$. Correspondingly, the
boundary conditions take the form:
\begin{equation}
 \label{bound_cond}
 \begin{array}{l}
 f(-W/2, \varphi)=0
 \:    ,  \;    - \pi/2 < \varphi < \pi/2 \:  ;
 \\
 f(W/2, \varphi)=0
  \:\: ,  \;  \;\;\;  \;  \pi/2 < \varphi < 3 \pi/2  \: .
 \end{array}
\end{equation}

{\em 3. The Ohmic and the hydrodynamic corrections
 to the ballistic conductance.}
The kinetic equation (\ref{kin_eq_with_gamma})
with the zero right part and the boundary conditions
(\ref{bound_cond}) is   easily solved. The result is
 $ f(y,\varphi) =   f_+(y,\varphi)$ at $  -\pi/2 < \varphi < \pi/2 $
and $  f(y,\varphi) =   f_-(y,\varphi) $ at
$ \pi/2 < \varphi < 3 \pi/2 $, where
\begin{equation}
  \label{f_pm}
f_{\pm} (y,\varphi)
 =
 E\,\frac{\sin\varphi  }{\gamma}
 \left\{
 1-\exp \left[
 \displaystyle
 -\frac{\gamma \, (
 y \pm W/2
 ) }{ \cos \varphi }
  \right]
 \right\}
 \:.
\end{equation}
In the ballistic regime, $\gamma W \ll 1$, for the angles $ \varphi$ being
not very close to $\pm \pi/2$ the equation~(\ref{f_pm})
  is simplified to
$ f_{\pm} (y,\varphi)  =  E\, (y \pm W/2) \, \tan \varphi  $.
This distribution describes the particles accelerating
 due to external field and colliding with the edges without
 any other scattering inside the bulk of the sample.

  We will use the truncated formula for
    the generalized  current density
$ j(y)= \int  _{0} ^{2\pi} d \varphi  \: \sin \varphi \, f(y,\varphi) $.
For a long sample,   $L \gg 1/\gamma$ the current density
corresponding  to Eq.~(\ref{f_pm}) at $\gamma W \ll  1$ is
\begin{equation}
 \label{j}
   j (y) =
2E \sum_{\pm}
\left( W/2  \pm y \right) \ln\left[
\frac{1}{ \displaystyle \gamma    \left( W/2  \pm y \right)    }
\right]
 \:.
\end{equation}
In the main order by $\ln\left[1/(\gamma W\right)]$
 this current density is homogeneous and equal to
$ j _0  =2EW\ln\left[1/(\gamma W)\right] $ [see Fig.~1(b)].
For the total current $ I= \int   _{-W/2} ^{W/2} dy  \, j(y) $
  we obtain from Eq.~(\ref{j}):
\begin{equation}
  \label{I_0}
I  = 2 EW^2 \Big[\ln\Big(\frac{1}{\gamma W}\Big)
  + \frac{1}{2}\, \Big]
\:.
\end{equation}
The logarithmic term in Eq.~(\ref{I_0}) is related
 to the particles moving along the trajectories with the
 angles $\varphi  \approx \pm \pi /2 $. A particle
 on such ``special'' trajectories  spends a longer time
between scattering events  on the opposite edges
as compared with the particles moving   along
the ``regular'' trajectories  with    $\varphi \sim 1 $
and,  thus, acquires a larger    velocity correction
due to acceleration by the field $ E $.

If the sample length is smaller than the total scattering length,
 $L \ll 1/\gamma$, than Eq.~(\ref{f_pm})
leads in the main order by $\ln(L/W)$ to the result:
\begin{equation}
  \label{I_0_L}
I = 2EW^2 \ln \Big( \frac{L}{W} \Big).
 \end{equation}
Possibly, the dependence~(\ref{I_0_L}) was observed
 in the work  \cite{exp_gr_4} in which
the thermal conductance of suspended graphene as
 a function of the sample length was measured.

The result~(\ref{I_0_L}) can be applied also
 to the limiting case of short samples, $L \sim W$,
or the samples having curved edges  with
a characteristic radius $R \sim W$. In these cases
 the scattering length is constrained just by
 the value $W$ and Eq.~(\ref{I_0_L}) yields:
 $I \sim EW^2 $ \cite{MI_pr}.

 In Fig.~1(c) we schematically  show
 the dependence of the total current $I$ on the
 scattering length  $l_N = 1/ \gamma_N $
 at a fixed value of the scattering length  $l_U=1/\gamma_U  \gg W$
  in the ballistic, $l_N \gg W$,
  as well as in the Ohmic, $l_N \ll W^2 /l_U$, and
  the hydrodynamic, $W^2 /l_U \ll l_N \ll W$, regimes.
  It is noteworthy that this dependence is non-monotonic
and saturates at very high as well as at very low values of $l_N$.

A more precise solution of
 Eq.~(\ref{kin_eq_with_gamma}) provides a hydrodynamic correction
 $\delta I _h$ to $I_0$. Such correction is related to
 the inter-particle normal collisions conserving momentum,
 which protect particles from a loss of their momentum in
  scattering on  edges.

In order to calculate the hydrodynamic correction in the ballistic limit $\gamma W \ll 1$,
the distribution function  should be presented in the form
$ f=f_0+f_1 $,  where $f_0$ is the function (\ref{f_pm})
 and $f_1$ is a correction to $f_0$ due to the right part
  of Eq.~(\ref{kin_eq_with_gamma}).
 The equation for $f_1$ at $\gamma_U= 0 $ takes the form:
\begin{equation}
 \label{Eq_f0_f1}
   \Big[\cos\varphi \, \frac{\partial }{ \partial y } + \gamma_N \Big] f_1
 = \gamma _N  P_{\sin} [f_0]
 \:,
\end{equation}
where $P_{\sin}$ is the projector  on the function $\sin \varphi$.
For its action on $f_0$  we have $P_{\sin}[ f_0] = j(y) \sin \varphi /\pi $,
  where $j(y)$ is given by Eq.~(\ref{j}). In the main order
  by the logarithm $\ln[1/(\gamma_N W)]$   the current density
  is homogeneous, $ j(y) \approx j_0  $,  therefore  the right part
of Eq.~(\ref{Eq_f0_f1})  becomes equal just to $ q E \sin \varphi $,
    where $ q= ( 2 \gamma _N W /\pi)  \ln [1/(\gamma _N W)]   \ll 1$.
  By this way, Eq.~(\ref{Eq_f0_f1}) turns  into
  Eq.~(\ref{kin_eq_with_gamma})  with zero right part
  and the value $E$ replaced by $qE$. Thus for all
  the values related to the   first hydrodynamic
  correction $f_1$  we just have: \
  $ f_1 (y, \varphi) = qf_0 (y, \varphi) $
   and $ I_1  \equiv \delta I_h = qI_0 $, namely:
\begin{equation}
  \label{delta_I_H}
   \delta I _h = \frac{4 E \gamma _N W^3}{\pi}
    \ln\Big(\frac{1}{\gamma_N W}\Big)^2
 \:.
\end{equation}
 This positive correction is a {\em precursor }of forming
the Poiseuille flow of a viscous fluid
 related to the interparticle collisions conserving momentum.

{\em 4. The kinematic magnetoresistance. }
Now we  study the effect of weak magnetic
field perpendicular to the 2D layer  on heat  and
charge ballistic transport of charged particles (electrons).
  For better transparency, we do the consideration
  by the two ways: (i) calculation of the magnetic field
  corrections to the trajectories of individual electrons;
  (ii) solution of the kinetic equation
 with the magnetic field term treated as a perturbation.

(i) The correction to the velocity of an electron  due to
infinitely small electric field,  $E \to 0$, during
a fixed time interval, $0<t<t^*$,  is just $ \delta v _x^ E (t)  = Et $
 at enough  weak magnetic field.
 In order to calculate the total current, $I \sim E$,
  one needs to average $ \delta v _x^ E  (t)  $   over
  the electron motion between the opposite sample edges
and then to sum up the averaged by time corrections
 $ \langle \delta v _x^ E   \rangle   _t $ by  all electrons.

 For this purpose, we present   the total  current $I$
   as a sum   of  the infinitely small
   contributions $dI(\varphi)$ coming from  the electrons
  with the angles $\varphi$ of the initial velocities
in the intervals  $(\varphi , \varphi + d \varphi ) $:
$ I = \int _0 ^{2\pi} d I(\varphi) $.
In order to calculate $dI(\varphi)$, one needs to know
   the electric field correction
  $ \delta v _x^ E (t)$   at the times $0 \leq t\leq t^*$,
   where  $t=0$ corresponds to scattering
   of an electron on one of  the sample edges
   and $t=t^*(\varphi )$ is the time when the  electron
   reaches the opposite edge.
 As a result,
 we obtain  $d I (\varphi)  = W  E  \, t^*(\varphi)  \,  d \varphi /2$.
 For the electrons moving from the left
 to the right the equation for $t^*$ is [see Fig.~1(a)]:
 \begin{equation}
 \label{Eq_t_zv}
  y_{\pm}(t^*) = W/2
  \: .
  \end{equation}

Solution of Eq. (\ref{Eq_t_zv}) in the limit
 of a narrow ballistic sample, $\gamma W \ll 1$, and
 weak magnetic fields, $B \to 0$, and integration
 of $d I (\varphi)$ over $\varphi$ leads to the
  result \cite{SM}:
\begin{equation}
  \label{I_B_straight}
I = 2 E W ^2 \Big[\,  \ln\Big(\frac{1}{\delta_m}\Big)  +
 C \, \frac{\omega_c^2 W^2 }{\delta^4_m } \,\,  \Big]
 \:,
\end{equation}
where  $\omega_c$ is the cyclotron frequency;
 $\delta_m$ is the characteristic minimum value of the ratio  $|v_y|/|v_x|$
 for  the electrons which scatter {\em only} on
 the edges [see Fig.1(a)];
$C$ is a numeric constant depending on the exact value
of $\delta_m$,  which cannot be determined within
the approach (i).

 For the long samples, $L \gg 1/\gamma$,
 and in the absence of the Umklapp scattering, $\gamma_U=0$,
the parameter  $\delta_m$ is equal to $ W/ l_{N}= \gamma W$ and
Eq.~(\ref{MR_B_straight_gamma})   leads to the magnetoresistance
 \begin{equation}
  \label{MR_B_straight_gamma}
 \frac{R (B) - R(0) }{  R(0)}  \sim   - \frac{ \omega_c^2   }
 { \gamma^4 W^2 \ln[1/(\gamma W )]}
 \:, \;\;
 \omega_c \ll \gamma ^2 W
  \:.
  \end{equation}
 For not very long samples, $W  \ll L \ll 1/\gamma$,
 we have $\delta_ m = W/ L$ and thus
 \begin{equation}
  \label{MR_B_straight_L}
 \frac{R (B) - R(0) }{  R(0)}  \sim
  - \frac{ \omega_c^2 L^4  }{ W^2 \ln(L/W )}
 \:,
 \;\;\; \omega_c \ll W/L^2
 \:.
\end{equation}
In the limiting case when the  sample is short,
$L \sim W$, or the sample edges are curved with
 a characteristic radius $R \sim W$,  the trajectories
 with all angles $\varphi$ are equally important.
 Thus one should put  in Eq.~(\ref{I_B_straight})  the  parameter
 $\delta _m $ equal to  $1$. In this way, the total current
 at weak magnetic fields, $\omega_c W  \ll 1 $, is estimated as
$   I =I_0+ \delta I $, where $   I_0 \sim    E W^2  $,
$   \delta I \sim   E W^4 \omega_c^2    $.
 The corresponding magnetoresistance is:
\begin{equation}
  \label{MR_B_simplest}
 \frac{R(B) -R(0) }{R(0) } \sim  - \omega_c^2 W^2
  \:,
 \;\;\; \omega_c \ll 1/W \:.
\end{equation}

It is noteworthy that the obtained ballistic
magnetoresistance (\ref{MR_B_straight_gamma}),
(\ref{MR_B_straight_L}), and (\ref{MR_B_simplest})
is negative. This is related to an increase
 of the mean length of the electron trajectories between the edges [see Fig.~1(a)]
 For not very long samples, $L \ll 1/\gamma$, the magnetoresistance
 is independent of $\gamma$
and, thus, of temperature.

(ii) Magnetoresistance of long straight samples,
 $L \gg 1/\gamma$,   can be also derived
within a rigorous solution of the  kinetic equation.
In the  presence of enough weak magnetic field
 and at  $\gamma_U=0$ it is:
\begin{equation}
 \label{kin_eq_B_gen}
\cos(\varphi) \frac{\partial f}{ \partial y }   -
 \omega_c \frac{\partial f}{ \partial \varphi }  -  \sin(\varphi) E
 = \mathrm{St}_N [f]
 \:.
\end{equation}
 We will seek the solution of Eq.~(\ref{kin_eq_B_gen})
 as a series $ f=f_0+f_1+f_2 $, where $f_0$ is given
 by Eq.~(\ref{f_pm}), while $f_1$ and $ f_2$ are proportional
 to the powers of magnetic field:
 $ f_1\sim \omega_c $,  $ f_2 \sim  \omega_c^2 $.
 The first order correction   to the current,
 $I_1 \sim f_1$, vanishes due to the symmetry
 between the ``+'' and ``-''  trajectories [see Fig.1(a)].
 The second order correction  $I_2 \sim f_2$ is nonzero and
 predominantly comes, as $I_0$, from the particles
  with the velocity angles $\varphi  \approx \pm \pi/2 $.
The calculations of $f_2$  leads to the equation
 (\ref{I_B_straight})   for the total current $I$
 with $\delta_m = \gamma W  $ and  $C= 3/4 $
 (see Ref.~\cite{SM} for details of calculations).

{\em 5. Discussion and conclusion.}
In the high-mobility GaAs quantum wells the strong
negative magnetoresistance was observed at
 low temperatures \cite{exp_hydgr_1,exp_hydgr_1_2,exp_hydgr_1_3,exp_hydgr_1_4,exp_hydgr_1_ov}.
Often, but not always the experimental magnetoresistance curve
 consists of the two peaks:
 the narrow sharp temperature-independent peak
 in vicinity of zero magnetic field and
 the  wider  temperature-dependent  peak
 with a large amplitude.
 The analysis of the dependence of the wide peak on temperature
  allowed  to explain it as a manifestation of forming
 a viscous flow of 2D electrons   \cite{hydr_tr_th_9}.

  The small peak is  independent
 of temperature and has the halfwidth
 of the order of 50~Oe \cite{exp_hydgr_1,exp_hydgr_1_2,exp_hydgr_1_3,exp_hydgr_1_ov}.
 Such magnetic field corresponds to the
 cyclotron diameter 2$R_c$ equal to 30-40 $\mu$m
 for the experimental values  of the electron densities
 2$\cdot10^{11}$-3$\cdot10^{11}$cm$^{-2}$.
 These values of $2R_c$ are comparable
 within  the order of magnitude with the
 typical sample widths or with the distance
 between the macroscopic oval defects which
 are often present in the high-mobility
 GaAs structures \cite{exp_hydgr_1_ov}.
 The ballistic resistivity $\varrho = E W /I$
 in the limiting case $L \sim W $
 at $W = $40 $~\mu$m is about 5~$\Omega$.
  This gives an estimation of the amplitude
  of {\em kinematic  } magnetoresistivity,
   $\varrho |_{\omega_c=1/W} - \varrho |_{\omega_c=0}$,
   which corresponds by the order of magnitude to the experimental values of
   amplitude of the small peak \cite{exp_hydgr_1_ov}.

 In this way, we have the evidences that the magnetoresistance
  (\ref{MR_B_straight_L})  and  (\ref{MR_B_simplest}),
  predicted in this work for not very long samples,
   $L \ll 1/\gamma$,  could be observed in
   Refs.~\cite{exp_hydgr_1,exp_hydgr_1_2,exp_hydgr_1_3,exp_hydgr_1_ov} as a small peak
   independent of temperature.
 Simultaneous  manifestation of the negative temperature-dependent viscous
 magnetoresistance and the negative temperature-independent ballistic
 magnetoresistance can be related to the presence
 in a given sample  of the both narrow
 and wide conductive regions in which the viscous
    and the ballistic regimes can be mutually realized.

{\em 6. Acknowledgements.}
 We are grateful to A.~I.~Chugunov, A.~P.~Dmitriev, and
V.~Yu.~Kachorovskii  for valuable discussions as well as to
  A.~P.~Alekseeva, E.~G.~Alekseeva, I.~P.~Alekseeva,
  N.~S.~Averkiev, A.~I.~Chugunov,  P.~S.~Shternin,
  and D.~S.~Svinkin for advice and support.
 The part of this work devoted to the study of  the hydrodynamic
 and Ohmic corrections to the ballistic conductance
 (Sections 2 and 3) was supported by the Russian Fund for
Basic Research (Contracts No. 16-02-01166-a)
 and by the grant of the Basis Foundation;
 the part of this work devoted to the study
 of the effect of magnetic field on the ballistic conductance
 (Section 4) was supported by the Russian
Science Foundation (Grant No.  17-12-01182).

\appendix

\section{ Supplemental material  }
\section{ Analysis of the structure of the kinetic equation  and details of its solution }

\subsection{1.  Exact solution of the kinetic equation for a given space harmonic}
 In this section,  we will study
  the structure
  of the kinetic equation~(M3) and of its solutions
   in different limiting regimes:
  the Ohmic, the hydrodynamic, and the ballistic regimes.
 Here and further
  we use the notations (M1), (M2),  ... for
 the references on the formulas in the main text.
 In order to understand
 general properties
  of the kinetic equation and of its solutions, it is instructive to  obtain an
  exact solution of Eq.~(M3)
  for a given space harmonic,
  $f(y,\varphi)\sim e^{iky}$, without construction
  of the solution of Eq.~(M3) in real space for an exact boundary conditions.

  In the main text we assumed the field $E$
  to be homogenous. However,  in this section it is convenient to assume
  that the field $ E $
 has a nontrivial dependence on the transverse
 space coordinate $y$: $E=E(y)$.

  Following to Refs.~\cite{hydr_tr_th_10,hydr_tr_th_11},
  we introduce the Fourier decomposition
   of $E(y)$ and $f(\varphi, y)  $
 by the space coordinate $y$:
\begin{equation}
 \label{Fourier}
E(y) = \sum \limits _{k   } E_k  e^{i k y }
,\; \;\;
  f(\varphi, y)  = \sum \limits _{k   } f_k(\varphi ) e^{i k y } .
\end{equation}
For the general analysis of Eq.~(M3),
  performing in this Section,  we  do not need
  to specify the exact values of the wavevectors
  $  k   $ in the expansions (\ref{Fourier}).

For each harmonic $ f(y,\varphi) =f_k(\varphi)  e^{i k y } $
 the kinetic equation (M3)
  becomes algebraic by the variable $k$
   and integral by $\varphi$:
\begin{equation}
\label{kin_eq_Four_1}
\left[i k \cos(\varphi)   + \gamma \right] f_k - \sin(\varphi) E_k
 = \gamma_N P [f_k] + \gamma_U P_0 [f_k] .
\end{equation}
where the projector operators $P$ and $P_0$
has the form: $P=P_{-1}+P_0+P_1$,
\begin{equation}
P_{m} [f] (\varphi)
 = e^{i m \varphi} \int \limits_{0} ^{2\pi} \frac{d \varphi' }{2\pi } \, e^{ - i m' \varphi} f(\varphi')
\: ,  \; \; m = \pm  1,0\, .
\end{equation}

The projector operators satisfy to the relations:
$\sin \varphi  = P \sin \varphi$, $P_0 = P P_0$,
 $P= P^2 $.  Using them, we can rewrite
  the kinetic equation in the form:
\begin{equation}
 \label{kin_eq_Pf}
P f_k =  E_k g \sin \varphi + g \left(\gamma_N P [f_k] +\gamma_U P_0 [f_k]  \right) \:,
\end{equation}
where $ g=PK^{-1}P $, $ K= i k \cos \varphi   + \gamma$.
The operator $g$ actually acts in the space of the harmonics
 $e^{im \varphi}$ with  $m=1,0,-1$.
That is, $g = g(k)$ is a $3 \times 3$
 matrix with the elements:
\begin{equation}
g_{mm'}= \int \limits_{0} ^{2\pi} d\varphi'
\frac{e^{i \varphi (-m+m')}}{2\pi}
\frac{1}{ik\cos\varphi + \gamma}
\:.
\end{equation}
In the  exact form we have:
\begin{equation}
g= \left(\begin{array}{ccc}
g_0 & g_1 & g_2
\\
g_1 & g_0 & g_1
\\
g_2 & g_1 & g_0
\end{array}\right)
\:,
\end{equation}
where
\begin{equation}
g_0 = \frac{1}{\gamma} \frac{1}{\sqrt{1+(k/\gamma)^2 } }
\:,
\end{equation}
\begin{equation}
g_1 = - \frac{i}{k} \Big[ 1- \frac{1}{\sqrt{1+(k/\gamma)^2 } } \Big]
\:,
\end{equation}
and
\begin{equation}
g_2 = - \frac{ \gamma }{k^2} \left[  \sqrt{1+(k/\gamma)^2 }  +
\frac{1}{\sqrt{1+(k/\gamma)^2 } }
-2 \right].
\end{equation}

The function $F(\varphi) =P [f_k]  (\varphi)$,
can be represented by the vector $F= (F_1, \, F_0, \, F_{-1})$
  consisting of the coordinates of $ F(\varphi)$ in the basis
 $e^{  i m \varphi}  $,   $m = 1,0,-1$. The kinetic equation
 (\ref{kin_eq_Pf})  yields for $F$
 the finite-dimensional equation containing
 the $3 \times 3$ matrixes $g$ and
$r=\mathrm{diag} ( \gamma_N , \gamma , \gamma_N )$:
\begin{equation}
\label{kin_eq_matrix}
F = \frac{E_k}{2i} g   \left(\begin{array}{c} 1 \\0 \\ -1 \end{array} \right)
+ grF
\:.
\end{equation}

The solution of the kinetic equation (\ref{kin_eq_Pf})
 can be easily performed. For $F$ we have from
 Eq.~(\ref{kin_eq_matrix}):
\begin{equation}
\label{F}
 F= \frac{E_k}{2i} (I-gr)^{-1}  g \left(\begin{array}{c} 1 \\0 \\ -1 \end{array} \right)
\:,
\end{equation}
where $I$ is the $3 \times 3$ unit matrix.
 Now one can easily find from Eq.~(\ref{kin_eq_Four_1})
  all angular harmonics of the distribution function
   $ f_k(\varphi) $:
\begin{equation}
\label{f_K_predv}
f_k(\varphi) =
 \frac{  \left\{ \gamma_N P [f_k] + \gamma_U P_0[ f_k ] \right\} (\varphi )+ E_k \sin \varphi  }{ ik\cos \varphi + \gamma}
 \:.
\end{equation}
Using Eq.~(\ref{F}) for $P [f_k]$ and $P_0[ f_k ]$,
 we obtain from Eq.~(\ref{f_K_predv}) the final form of
 the relation between $f_k(\varphi)$ and $E_k$:
\begin{equation}
\label{fCE}
f_k(\varphi) = C(k,\varphi) E_k \:,
\end{equation}
where $C=C(k,\varphi)$ is expressed via
 the matrixes $r$ and $g=g(k)$ as:
\begin{equation}
\label{C}
 C  =
\frac{
 \left(\begin{array}{c} e^{i \varphi} \\ 1 \\ e^{ - i \varphi} \end{array} \right) ^ T
[r (I-gr)^{-1} g +I]
  \left(\begin{array}{c} 1 \\0 \\ -1 \end{array} \right)
 }
{ 2i \,( \, ik\cos \varphi + \gamma \,) }
 .
\end{equation}

  First, the obtained general result (\ref{C})
 shows that the zero ($m = 0 $) harmonic
 in the distribution function   $f(y,\varphi)$
 is equal to zero: $f_{mk} =0$ at any  $k$.
This means that no perturbation of the particle
density arises  in the linear response
on  the external field $E$.

Second, let us consider the different
 limiting regimes of transport on base of Eq.~(\ref{C}).

 In order to consider the simplest Ohmic regime,
 let us again suppose that  the external field  is  homogeneous, $E(y) = const $,
 and it is possible not to impose any boundary conditions
 on $f(y,\varphi)$ on the longitudinal sample boundaries
 $y= \pm W/2$. For the zero space harmonic,
  $k=0$, the matrix $I-gr$ is degenerated. The kernel of the matrix $I-gr$
  contains the functions $f(\varphi) = const$ corresponding
  to perturbation of the particle density,
  which is conserved in our model.
According to this fact,
 for the distribution function which is homogeneous
 and does not contain a perturbation of the particle density
 we have from Eq.~(\ref{kin_eq_matrix}):
\begin{equation}
\label{Ohm}
f(y, \varphi) = \frac{E}{\gamma_U} \sin  \varphi
\:.
\end{equation}
Herewith, the inter-particle scattering
conserving momentum does not play any role.
 Equation~(\ref{Ohm}) describes the
 Ohmic regime of transport. For the total current
 in this simplest regime we just obtain from Eq.~(\ref{Ohm})
  the usual Drude formula:
  \begin{equation}
  I = \frac{ \pi E W }{\gamma_U  }
  \: .
  \end{equation}

If the actual wavevectors  $k$
are small as compared with the total scattering
length, $k \ll 1/\gamma$, one should expect
 that  Eq.~(\ref{C}) describes the
hydrodynamic regime of transport. Herewith
 let
 us consider that
 the normal scattering conserving momentum
 is much more intensive
than scattering not conserving momentum:
$\gamma_N \gg \gamma_U$.

Indeed, in the case $k \ll 1/\gamma$ the elements
 of the matrix $g$ have the asymptotes:
\begin{equation}
\begin{array}{c}
 \displaystyle
g_0 (k)  = \frac{1}{\gamma}  -  \frac{1}{2} \frac{k^2}{\gamma^3}
 \: ,
\\
\\
 \displaystyle
g_1 (k) =-i\frac{k}{2\gamma^2}
 \: ,
  \;\;\; \;\;
g_2 (k) = -  \frac{1}{4} \frac{k^2}{\gamma^3}
\:.
\end{array}
\end{equation}
The matrix in  Eq.~(\ref{C}) becomes small
 as compared with unity:
\begin{equation}
   \label{norm}
 || r (I-gr)^{-1} g +I   || \ll 1
 \:,
 \end{equation}
 where $||.||$ is some matrix norm.
 Equation~(\ref{C}) for the coefficients $C$ gives:
\begin{equation}
 \label{H}
C(k,\varphi)
=
 \frac{\sin \varphi }{\gamma_U + \gamma_N (k/2\gamma)^2 }
 \:.
\end{equation}
Equations
 (\ref{fCE}) and (\ref{H})
 after the Fourier transform by $k$
 and integration over   the variable $\varphi $
 with the factor $\sin \varphi  $
 lead  to the Navier-Stocks equation
 for the current density $j(y)$:
\begin{equation}
 \label{Navier}
\frac{1}{4 \gamma_N } j'' -  \gamma_U j +E (y)=0
\:.
\end{equation}
Here $1/(4 \gamma_N  )$ is the viscosity coefficient.
As it is well-known \cite{book__gas__the_method},
in this case only the first and the second angular harmonics
of the distribution function are important
for describing the flow and deriving
  the Navier-Stocks equation (\ref{Navier}).
For the total current $I $ at $\gamma_U=0$ we have
\begin{equation}
I = c_h E \gamma_N W^3
\: ,
\end{equation}
where $c_h$ is a numeric constant.

One can see from Eqs.~(\ref{C}) and (\ref{norm})
 that in the hydrodynamic regime, $\gamma W \gg 1$,
the left and the right parts of Eq.~(M3)
 are of the same order of magnitude and both
  are important for deriving the corresponding
  Navier-Stocks equation  (\ref{Navier}).

Note that the above consideration
of the hydrodynamic regime does not describe
 the near-edge regions of the widths of the order of $1/\gamma$,
 where the profile of the current density
 is determined by both inter-particle scattering
 and scattering on the sample edges and
  the hydrodynamic boundary conditions,
   $j(y=\pm W/2 )  = 0 $, are formed.

For the  narrow samples, $\gamma W \ll 1 $, or
for the near-edge regions  of wide samples,
the flow is ballistic.   The profile of the distribution function
and the current density   is formed by the  space harmonics
 with the large wavevectors $k \sim 1/W$, $k \gg 1/\gamma$.
Correspondingly, one needs to take into account many harmonics
 by the variable $\varphi$ in the distribution function  $f_k(\varphi)$
 in order to describe
 the flow \cite{book__gas__the_method}.

In the case  $k \gg 1/\gamma$  the elements of the matrix $g$
have the asymptotes:
\begin{equation}
g_0 (k) = \frac{1}{k}
, \;\;\; \;\;
g_1 (k) =-\frac{i}{k}
, \;\;\; \;\;
g_2  (k) = -   \frac{1}{k}
\:.
\end{equation}
For the matrix in Eq.~(\ref{C}) we obtain the estimations:
\begin{equation}
r (I-gr)^{-1} g +I \approx I
\end{equation}
and
\begin{equation}
||r (I-gr)^{-1} g ||\sim \frac{ \gamma}{k}
\:.
\end{equation}
As a result, the main part of the coefficient $C$ is
\begin{equation}
C(k,\varphi)
=
  \frac{\sin \varphi }{ ik \cos \varphi   + \gamma }
  \:,
\end{equation}
and the corrections to this expression,
 which are proportional to the right part of  Eq.~(M3),
 have the relative magnitude of the order of $\gamma/k$.

In this way,  in the ballistic regime, $\gamma W \ll 1$, the right part of  Eq.~(M3)
 is far smaller than the left part and the flow profile
 in the main order by $\gamma / k $ should be calculated
 from the terms in the left part of Eq.~(M3).

\subsection{2.  Details of calculations of magnetoresistance }
In this section, we will present some details of the calculation
 of magnetoresistance in the ballistic regime
 by the two methods described in the main text:
  by analysis of the corrections
  to the trajectories of individual electrons
  [method (i)]
  and by solving the kinetic equation
  with the magnetic field term
  being a perturbation
  [method (ii)].

   (i) If the sample is long, $L \gg W $, and its edges are straight, but rough,
the main contribution to the total current comes
 from the electrons with
 the initial velocity angles $\varphi \approx \pm \pi /2$ (see Section 3 of the main text).
For such electrons  the initial conditions
on  their  trajectories  $\mathbf{r}_{\pm}(t)$ at $t=0$ are:
 \begin{equation}
 \label{left}
 \begin{array}{l}
 \displaystyle
 x=0 \: , \qquad y=-W/2 \: ,
 \\
 \displaystyle
  v_x = \pm 1
  \:, \qquad
  v_y=\delta >0 \:,
  \end{array}
  \end{equation}
  or
   \begin{equation}
 \begin{array}{l}
 \displaystyle
 x=0 \: , \qquad y=W/2 \: ,
 \\
 \displaystyle
  v_x = \pm 1
  \:, \qquad
  v_y=\delta < 0 \:,
  \end{array}
  \end{equation}
   where the parameter $\delta $ is small:
  $ |\delta | \ll 1$ [see Fig.~1(a) of the main text].
  The angles of the initial  velocities
  are $\varphi \approx \pm (\pi/2 - \delta )$,
  where $|\delta| \ll 1 $.
 Correspondingly, for the total current we have:
 \begin{equation}
  I \approx  4 \int  _{\delta_m} ^{1} dI(\delta )
  \:.
\end{equation}

      There exists the characteristic minimal  possible value
      of the absolute value of the parameter $\delta $ of the electrons which scatter
      only on the sample edges: $\delta _m  \ll 1$ .
  If the sample is longer
 than the particle mean free path, $L \gg 1/\gamma $,
the angles of the trajectories with the directions close to the $x$ axis are constrained
by scattering and $\delta_m = \gamma W $. Otherwise, for not very long samples,
 $ W \ll L \ll 1/\gamma $, the maximum value of the parameter $\delta$
  is determined by the sample length:  $\delta_m =  W/L $.

   The solution of the Newton equation
 for an electron in magnetic field
   with the initial conditions (\ref{left}) is:
 \begin{equation}
  \left\{ \begin{array}{l}
x(t) =  \pm A \, [ \cos(\omega_c t + \phi_0) - \cos(\phi_0)]
 \\
y(t) =  - W/2+ A \, [ \sin(\omega_c t + \phi_0) - \sin (\phi_0)]
\end{array}
\right.
 \:,
\end{equation}
 where $ A = \sqrt{1+\delta^2} /\omega_c  $,  $ \phi_0 = \mp \mathrm{atan}( 1/\delta ) $.

The equation (M11) on the value $t^*$ in an explicit form is:
\begin{equation}
 W\omega_c = \delta \sin (\omega_c t ) \pm  [1- \cos(\omega_ct)]
  \:.
\end{equation}
Solution of this equation at $\omega_c \ll  \delta ^2 / W  $ should be performed
  by the perturbation theory by the small parameter
  $\omega_c W / \delta^2 \ll 1$. After some calculations we arrive to the result
  \begin{equation}
  t^*(\delta ) = \frac{W}{ \delta }
   \mp \frac{ \omega_c W^2 }{ 2 \delta ^3 } +
   \frac{ \omega_c ^2 W^3 }{ 2 \delta ^5 }
    \:,
\end{equation}
Thus the time $t^*$  is written as the series:
$ t^* = t^*_0  + t_1^* + t^*_2 $, where  $ t_0^* = W/\delta $
is the time $t^*$ in zero magnetic field, while $t_1^* $ and $ t^*_2$
 are the fist and the second order magnetic field corrections:
$ t_1^*  = \mp \omega_c W^2 /(2 \delta ^3) $,
$ t_2^*  = \omega_c^2 W^3 / ( 2 \delta ^5) $.
These expressions were used in the main text
 in order to calculate the total current $I$
 by integrating over the angle $\varphi$ (i. e., by the variable $\delta$).

(ii) In the kinetic equation~(M16)
 the part $\gamma _N P[f]$  of the collision integral $\mathrm{St}_N[f]$
 in the main order  by the logarithm $ \ln[1/(\gamma W) ] $ should
 be neglected (see the previous Section). Therefore the equation~(M17)
 takes the form:
\begin{equation}
\label{kin_eq_B_main}
\left[\cos(\varphi) \frac{\partial }{ \partial y } + \gamma \right] f -  \sin(\varphi) E
 = \omega_c \frac{\partial f}{ \partial \varphi }
 \:,
\end{equation}
  where the right part is a small perturbation to the left part.

We seek the solution of Eq.~(\ref{kin_eq_B_main})
 as the series $f=f_0+f_1+f_2$, where $f_n \sim \omega _ c ^n$, $n=0,1,2$.
 The function $f_0$ is given by Eq.~(M5), while $f_1$ and $f_2$
  satisfy to the equations:
 \begin{equation}
 \label{B_f_1}
 \left[\cos(\varphi) \frac{\partial }{ \partial y } + \gamma \right] f_1
  = \omega_c \frac{\partial f_0 }{ \partial \varphi }
 \end{equation}
and
 \begin{equation}
 \label{B_f_2}
 \left[\cos(\varphi) \frac{\partial }{ \partial y } + \gamma \right] f_2
  = \omega_c \frac{\partial f_1 }{ \partial \varphi }
  \:.
\end{equation}

Below we perform a calculation
 of the distribution function corrections
$f_1$ and $f_2$ from Eqs.~(\ref{B_f_1}) and (\ref{B_f_2})
in the interval of angles  $ 0 < \varphi < \pi/2$
 [see the trajectory $\mathbf{r}_{+}(t)$ on the Fig.~1(a) in the main text].
  We will also find the contribution
  to the current from the electrons  with these angles.
   The expressions for the distribution function and the contributions to the current  from
   the electrons with the rest   angles $\varphi$, $\pi/2 < \varphi < 2 \pi$, are similar.
   Therefore we will present only the final result for them
 without details of calculations.

As it had been discussed above,  the main contribution
 to the current from the electrons with the angles
 in  the interval $ 0 < \varphi < \pi/2$
 comes from the angles $\varphi =  \pi/2 - \delta$,
 where   $0< \delta \ll 1 $.
  For these angles we have:
\begin{equation}
\label{cos}
\cos \varphi \approx \delta
 \:.
\end{equation}
 The  expressions
 for the main part of the distribution function (M5)
and its derivative by $\varphi$ for these angles $ 0 < \varphi < \pi/2$,
 $\varphi \approx \pi/2$ take the forms:
 \begin{equation}
  \label{f0}
f_0 ^{+} (y,\delta) = \frac{E}{\gamma } \left[ 1-e^{
 \displaystyle
- \frac{\gamma}{\delta} \left(y + \frac{W}{2}  \right) }  \right]
 \end{equation}
 and
 \begin{equation}
 \label{right_part}
\frac{\partial f_0^{+} (y,\delta)}{ \partial \varphi }
=  \frac{E}{\delta^2}  \: \left(y + \frac{W}{2} \right) \:
 e^{ \displaystyle \displaystyle - \frac{\gamma}{\delta}  \left(y + \frac{W}{2} \right) }
 \:.
 \end{equation}

The solution of Eq.~(\ref{B_f_1})
with the right part (\ref{right_part}) is:
 \begin{equation}
f_1^{+} (y,\delta) = \frac{\omega_c E }{2 \delta ^2 } \left(y + \frac{W}{2} \right)   ^2
\: e^{ \displaystyle - \frac{\gamma}{\delta} \left(y + \frac{W}{2} \right) }
\:.
 \end{equation}
Substitution of this formula to the right part of
 Eq.~(\ref{B_f_2})  and solving the resulting equation yields:
 \begin{equation}
 \label{f_2}
\begin{array}{c}
\displaystyle
f_2^{+}(y,\delta) = \frac{\omega_c ^2  E }{ 2\delta^5 } \left(y + \frac{W}{2} \right) ^3
\times
\\
\\
\displaystyle
\times \left[ 1 -\frac{1}{4} \left(y + \frac{W}{2} \right)
 \frac{\gamma}{\delta} \right]
 \: e^{ \displaystyle - \frac{\gamma}{\delta} \left(y + \frac{W}{2} \right) }
 \:.
 \end{array}
 \end{equation}

For the magnetic field correction $I_2 ^{ \mathrm{I} }  $
to the total current $I$ from the particles
 with the angles $0 < \varphi < \pi/2$ we have:
 \begin{equation}
   \label{j_2__tot}
I_2^{ \mathrm{I} }  \approx  \int \limits _0^1 d\delta
                             \int \limits _{-W/2}^{W/2}
 d y \:  f_2 ^{+} (y,\delta)
\:.
 \end{equation}
 The second order contributions
 $I_2^{ \mathrm{II} }  $; $ I_2^{ \mathrm{III} } (y)  $; and $  I_2^{ \mathrm{IV} } $
 from the other diapasons of angles,
 $\pi/2 < \varphi < \pi $;  $\pi < \varphi < 2\pi/2 $; and  $3 \pi/2 < \varphi < 2\pi $,
 are the same as due to symmetry of the trajectories with $v_x = \pm |v_x|$ and $ v_y = \pm |v_y|$.
 Thus  the total magnetic field dependent part
 of the current is $ I_2 = 4 I_2^{ \mathrm{I} }  $.

  A calculation by Eqs.~(\ref{f0}), (\ref{f_2}), and (\ref{j_2__tot}) yields the final result
  for the total current in the zero and the second orders
   by the magnetic field $B \sim \omega_c \to 0$:
  \begin{equation}
 I = 2 E W^2 \left[   \ln\left(\frac{1}{\gamma W }\right)  +
  \frac{3 \omega_c^2  }{ 4W^2 \gamma^4 }  \right]
  \:.
 \end{equation}


\begin{thebibliography}{}
\bibitem{SM}
For the analysis of the structure of kinetic equation  in differnt regimes
and the details of its solution in the presence of magnetic field
 see Supplemetal Material.
\end{thebibliography}
\end{document}